\documentclass[12pt,preprint]{aastex}
\usepackage{graphicx}
\usepackage{txfonts}
\usepackage{amssymb}

\usepackage{vmargin}
\usepackage{array}
\newcommand{\bu}{\mathbf{u}}
\renewcommand{\bv}{\mathbf{v}}
\newcommand{\BZ}{\mathbf{B}_0}
\newcommand{\xh}{\hat{\mathbf{x}}}
\newcommand{\yh}{\hat{\mathbf{y}}}
\newcommand{\zh}{\hat{\mathbf{z}}}

\newcommand{\bB}{\mathbf{B}}
\newcommand{\bC}{\mathbf{C}}

\newcommand{\bx}{\mathbf{x}}

\newcommand{\bX}{\mathbf{X}}
\renewcommand{\div}{\mathbf{\nabla}\cdot}

\newcommand{\dxdy}[2]{{\frac{\partial{#1}}{\partial{#2}}}}

\newcommand{\DxDy}

\def\rhobar{{\bar\rho}}

\def\Tbar{{\overline T}}

\def\gbar{{\bar g}}

\def\nubar{{\bar\nu}}

\def\rhat{{\hat{\bf r}}}
\def\omegavec{{\bf \Omega}}
\def\grad{{\bf \nabla}}
\def\curl{{\bf \nabla} \times}
\def\vvec{{\bf v}}

\def\Bvec{{\bf B}}

\def\Jvec{{\bf J}}

\begin{document}
%\title[Observational signatures of IGW in stars]{Observational signatures of
 % internal gravity waves in massive stars}
\title{The hottest hot Jupiters may host atmospheric dynamos}
\author{T.M. Rogers}
\affil{Department of Mathematics and Statistics, Newcastle University,
  UK}
\affil{Planetary Science Institute, Tucson, AZ 85721, USA}
\author{J.N.McElwaine}
\affil{Department of Earth Sciences, Durham University, UK}
\affil{Planetary Science Institute, Tucson, AZ 85721, USA}

\begin{abstract}
Hot Jupiters have proven themselves to be a rich class of exoplanets
which test our theories of planetary evolution and atmospheric
dynamics under extreme conditions. Here, we present three-dimensional
magnetohydrodynamic simulations and analytic results which demonstrate
that a dynamo can be maintained in the thin, stably-stratified
atmosphere of a hot Jupiter, independent of the presumed
deep-seated dynamo. This dynamo is maintained by conductivity
variations arising from strong asymmetric heating from the planets'
host star. The presence of a dynamo significantly increases the
surface magnetic field strength and alters the overall planetary
magnetic field geometry, possibly affecting star-planet magnetic
interactions. 

%Dynamos in stably-stratified regions are largely non-existent because
%the ``alpha'' effect, which regenerates the poloidal field from
%toroidal field, is strongly resisted by the buoyancy force. This is
%particularly true in thin atmospheres where flows are largely
%two-dimensional and magnetic field generation runs into the well
%known Cowling anti-dynamo theorem. Here, we present
%three-dimensional dynamo simulations which demonstrate that a dynamo
%can be maintained in the thin, stably-stratified atmosphere of a hot
%Jupiter. This dynamo is maintained by conductivity variations
%arising from strong asymmetric heating from the planets' host star.
%This is the first instance of a variable conductivity driven dynamo
%in an astrophysical setting where one would not exist otherwise. The
%presence of a dynamo increases Ohmic heating in the deep atmosphere
%by () over models with a magnetic field permeating through the
%atmosphere from the presumed deep-seated dynamo. Such an atmospheric
%dynamo may provide the heating required to inflate these planets and
%would affect the planetary field morphology and hence, star-planet
%magnetic interactions.

\end{abstract}

\section{Introduction}
To date more than 5000 exoplanets have been discovered and a couple
hundred are considered ``hot Jupiters'' -- Jupiter sized planets close
to their host star. Hot Jupiters were the first detected exoplanets
and remain the best characterized due to their favorable observing
conditions. Because of their close proximity to their host star,
these planets are tidally locked with a constant day and nightside.
This asymmetric heating leads to strong eastward directed atmospheric
winds which have been studied extensively
\citep{cho03,show02,coop05,iandd08,show09,rm10,lewis10,tcho10,heng11,kataria16}.
While atmospheric dynamic calculations generally yield similar
results, such as eastward winds in excess of a km/s, observations
indicate varying circulation efficiency. Infrared observations have
demonstrated that hot Jupiters have a range of day-night temperature
differentials and there is some indication that this variation is
temperature dependent, with hotter planets showing larger
differentials than cooler planets \citep{cowan11,ks16}. 

Intense irradiation from the host star can lead to thermal ionization
of several alkali metals \citep{perna10a,bat10}. Therefore, hot
Jupiters are partially ionized. Numerous authors have
demonstrated that this ionization allows atmospheric winds to
couple to the deep-seated, dynamo driven magnetic field
\citep{perna10a,perna10b,bat10,menou12}. This coupling could lead to
currents which penetrate into the deep atmosphere, generating Ohmic
heating, which could in turn, contribute to the inflated radii observed in
half of all hot Jupiters \citep{bat11,wu13,gs15,gs16}. Magnetic
interaction could also reduce circulation
efficiency, particularly in hot planets where the day-night flow could
be impeded by the Lorentz force. These results demonstrate that
magnetism in hot Jupiters could have important observational consequences and thus,
warrant further investigation.

\cite{rs14} carried out the fist magnetohydrodynamic (MHD) simulations
of a hot Jupiter which self-consistently included Ohmic heating.
Those simulations found that inclusion of magnetic fields could severely affect the
atmospheric flows leading to variable and reversed winds. They also found that while
the MHD simulations did reproduce the qualitative picture proposed by
earlier theoretical work \citep{menou12,rm13}, they failed to
reproduce the amplitude of Ohmic heating required to explain inflated
radii \citep{rs14,rk14}. The discrepancy between 
theoretical models and numerical simulations leaves the viability of
the Ohmic mechansim for inflating exoplanets still in question.

Hot Jupiters also likely interact with their host stars' magnetic
field, possibly leading to observable features such as asymmetry in
the light curves of transiting planets \citep{vidotto10b,cauley15} and
induced activity in the atmosphere of their host star
\citep{shkolnik03,shkolnik05,shkolnik08}. Such interactions depend on
the planetary magnetic field strength and geometry
\citep{cuntz00,ip04}. Therefore, understanding the planetary magnetic field
is important if we are to correctly interpret such observations.

The day-night temperature differential on hot Jupiters leads to severe
day-night variations in ionization and hence, conductivity.
Similarly, there are large variations in conductivity between deep and
shallow atmospheric layers. \cite{bw92} showed that
variations in conductivity in the direction of the dominant flow,
could lead to a dynamo. More recently, \cite{pet16} showed that a
temperature dependent conductivity could produce a dynamo, even with
small temperature fluctuations and a weakly temperature-dependent
conductivity. Hot Jupiter atmospheres are perhaps the most asymmetric
astrophysical objects, with perhaps the largest temperature
(conductivity) variations and so provide an ideal testbed of the
theories outlined in those works. Here we present three-dimensional (3D) numerical simulations and
analytic results which show that a variable conductivity dynamo (VCD)
may proceed in some hot Jupiter atmospheres. 

\section{Numerical simulations of atmospheric dynamos}
We solve the full magnetohydrodynamic (MHD) equations in 3D in the
anelastic approximation, as described in \cite{rk14}. The model
solves the following equations:
\begin{eqnarray}
  \div \rhobar \vvec &=& 0, \label{eq:rho}\\
  \div \bB &=& 0,  \label{eq:sol} \\
  \rhobar \frac{\partial \bv}{\partial t}+\div(\rhobar \bv\bv) 
    &=& - \grad p - \rho \gbar \rhat  
        + 2 \rhobar \vvec \times \omegavec + \ldots\nonumber\\
    &&\mbox{}       
       + \div\left[2 \rhobar \nubar \left(e_{ij} - \frac{1}{3}\delta_{ij}\, ( \div \bv )\right)\right]
       - \frac{1}{\mu_0} \bB \times  ( \curl \bB),\label{eq:mom}  \\
   \dxdy{T}{t}+(\bv\cdot\nabla){T}
   &=&-v_{r}\left[\dxdy{\overline{T}}{r}-(\gamma-1)\overline{T}h_{\rho}\right]
       +(\gamma-1)Th_{\rho}v_{r} + \ldots \nonumber\\
   &&\mbox{}+ \gamma\overline{\kappa}\left[\nabla^{2}T+(h_{\rho}+h_{\kappa})\dxdy{T}{r}\right] 
      + \frac{T_{eq}-T}{\tau_{rad}}
      +\frac{\eta}{\mu_{o}\rho c_{p}}|\nabla \times \bB|^{2}. \label{eq:T}
\end{eqnarray}
Equation~\ref{eq:rho} represents the continuity equation in the
anelastic approximation \citep{go69,rg05}. This approximation allows
some level of compressibility by allowing variation of the reference
state density, $\rhobar$, which varies in this model by four orders of
magnitude. Equation~\ref{eq:sol} represents the conservation of
magnetic flux. Equation~\ref{eq:mom} represents conservation of
momentum including Coriolis and Lorentz forces. Equation~\ref{eq:T}
represents the energy equation including a forcing term to mimic
stellar insolation (fourth term on right hand side) and Ohmic heating
(fifth term on right hand side, where T$_{eq}$ is defined in
Equation~\ref{eq:temp}). All variables take their usual meaning and
details can be found in \cite{rk14}.

In the work presented here, the magnetic diffusivity $\eta$ (inverse
conductivity) is a function of all space. Therefore, the magnetic
induction equation is
\begin{eqnarray}\label{eq:induction}
  {\partial \Bvec \over \partial t}
  =\nabla\times\left(\vvec\times\Bvec\right)+\eta\nabla^2\Bvec-\left(\nabla\eta\right)\times\left(\nabla\times\Bvec\right).
\end{eqnarray}
  In the hot Jupiter system toroidal field can be generated from
  poloidal field by radial shear due to stronger winds at the
  planetary surface. Although the dynamo mechanism by conductivity variations is
  subtle, one can show that given the correct alignment between
  $\nabla\times\Bvec$ and $\nabla \eta$ the last term on the right hand
  side (RHS) of Equation 5 can provide a
  positive $\alpha$ effect, thus regenerating poloidal field from
  toroidal and closing the dynamo loop.

The magnetic diffusivity is calculated from the initial 
temperature profile given by: 
\begin{equation}\label{eq:temp}
  T_{eq}\left(r,\theta,\phi\right)=\Tbar(r)+\Delta T_{eq}(r),
  \cos\theta\cos\phi
\end{equation}
where $\Tbar(r)$ is the reference state temperature from 
\citep{rk14} and $\Delta T_{eq}$ is the specified day-night
temperature, which is extrapolated logarithmically from the surface to
10\,Bar. %Tami $10^6$\,Pa.
Using this temperature profile, the magnetic diffusivity is calculated
using the method from \cite{rm13} where:
\begin{equation}\label{eq:etaT}
  \eta\left(r,\theta,\phi\right)=230\frac{\sqrt{T_{eq}}}{\chi_e}
\end{equation}
and $\chi_e$ is the ionization fraction. The ionization fraction is
calculated at each point using the Saha equation taking into account all
elements from hydrogen to nickel and abundances from
\cite{Lodders10}. 

\begin{figure}[t]
\centering
\includegraphics[width=6in]{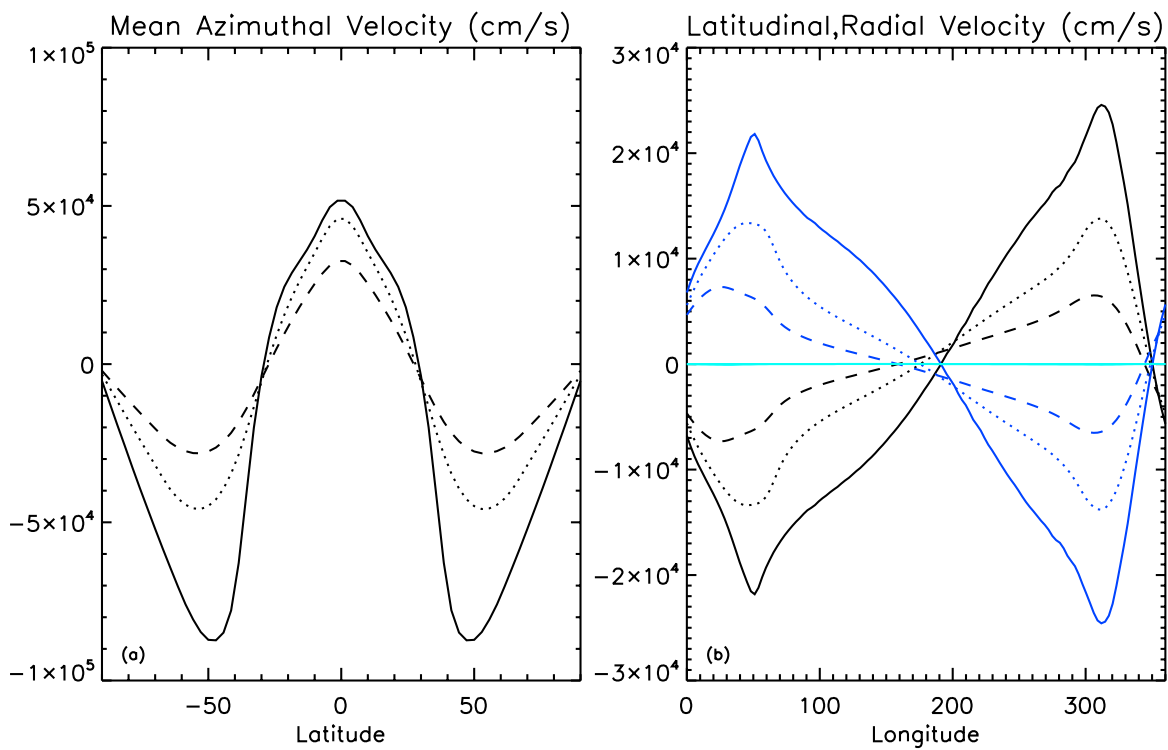}
\caption{Atmospheric winds in hot Jupiter atmosphere. (a) Time and
  longitudinally averaged longitudinal velocity at the surface (solid
  line), 95\% (dotted line) and 85\% of the computed domain (dashed
  line). (b) Latitudinal velocity averaged in time and over the
  Northern/Southern hemispheres (blue/black lines respectively) at the
  same depths as (a). Similarly averaged radial velocities are shown
  in cyan. (c) Time and longitudinally averaged velocity as a function
  of radius at the equator (solid line) and at mid-latitudes (dashed
  and dotted line).}
\label{flow}
\end{figure}
The fiducial numerical model we present is Model M8 from \cite{rk14},
but with $\Delta T_{eq}$=1000\,K, such that the night side at the top
of the domain is $\sim$1800\,K and the day side is $\sim$2800\,K. The
  atmospheric winds found in the hydrodynamic model are shown in
  Fig.~\ref{flow}. Near the surface, where the temperature forcing is
  strong, the model produces strong eastward directed jets at low
  latitudes, return flows at high latitudes and weaker, hemispheric
  meridional circulation. Deeper in the atmosphere, the forcing is
 reduced and winds fall off dramatically with
  depth. Radial flows are extremely weak throughout, with amplitudes
  0.1--1\% their horizontal counterparts. These winds are similar to
  those found in many other hydrodynamical simulations of hot Jupiters
  \citep{coop05}, with the main difference being that our winds are
  slightly weaker, probably due to the use of the full viscous term,
  rather than using a hyperdiffusivity. 

\begin{figure}
\centering
\includegraphics[width=6in]{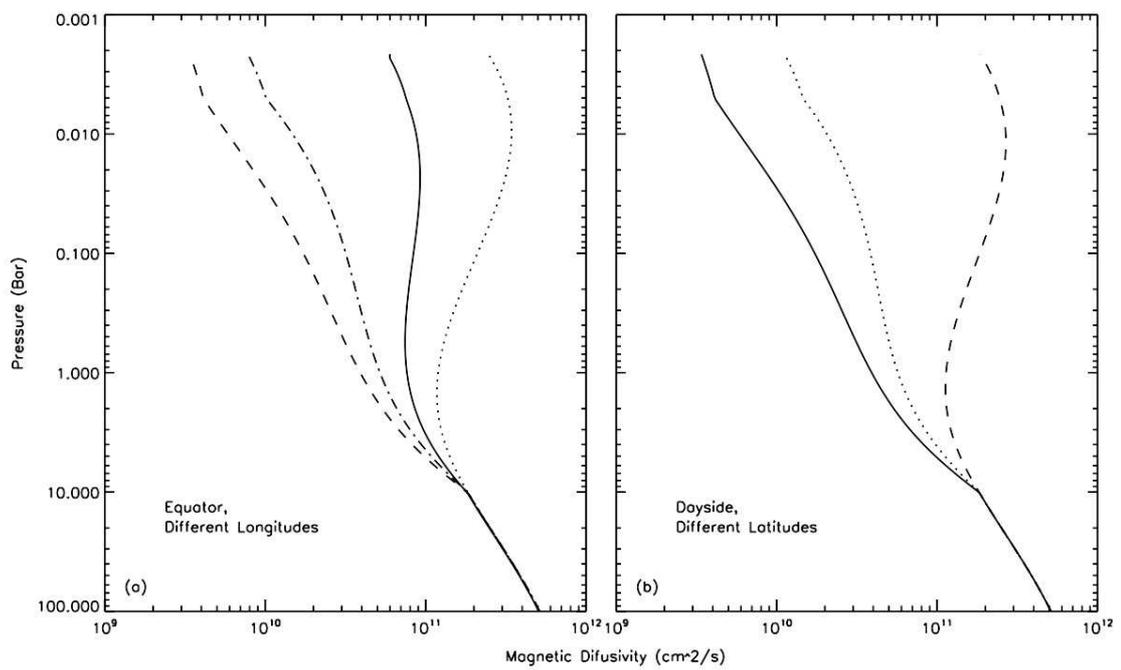}
\caption{Profile of magnetic diffusivity as a function of longitude
  (a) and latitude (b). (a) Radial profile of diffusivity at the
  equator at the sub-stellar point (dashed line), the nightside
  (dotted line), near the terminator (dash-dot line) and the mean
  (solid line). (b) Radial profile of diffusivity at the substellar
  point at the equator (solid line), mid-latitude (dotted line) and
  near the pole (dashed line).}
\label{eta}
\end{figure}
Magnetic effects are then investigated by including an initial
magnetic field of 5\,G at the bottom and 3\,G at the top of the
domain. Fig.~\ref{eta} shows the magnetic diffusivity as a function of
radius for various latitudes and longitudes for our model. This
``dynamo'' model has no continously imposed field, unlike the models
presented in \cite{rs14} and \cite{rk14}. However, to fully
investigate the effect of an atmospheric dynamo, we ran three
additional models: (1) ``constant $\eta$'' --- a model with a constant
magnetic diffusivity equal to the mean diffusivity
($5\times 10^{11}$), (2) ``imposed+dynamo'' --- a model with variable
conductivity, as shown in Fig.~\ref{eta} but with an imposed dipolar
magnetic field of strength 3\,G at the base of the simulated domain
meant to mimic the deep-seated, convectively driven dynamo and (3)
``imposed+constant'' --- a model with an imposed dipolar magnetic
field of strength 3\,G at the base of the simulated domain, but with a
constant magnetic diffusivity ($5\times 10^{11}$).

The magnetic diffusivity is \textit{not} a function of time in any of
the models. That is, it does not change due to advection of heat or
Ohmic heating. We will include this effect in forthcoming papers, but
discuss the possible relevance of a fully temperature-dependent
conductivity in Section 5.

\section{Numerical Results}

\begin{figure}
\centering
\includegraphics[width=6in]{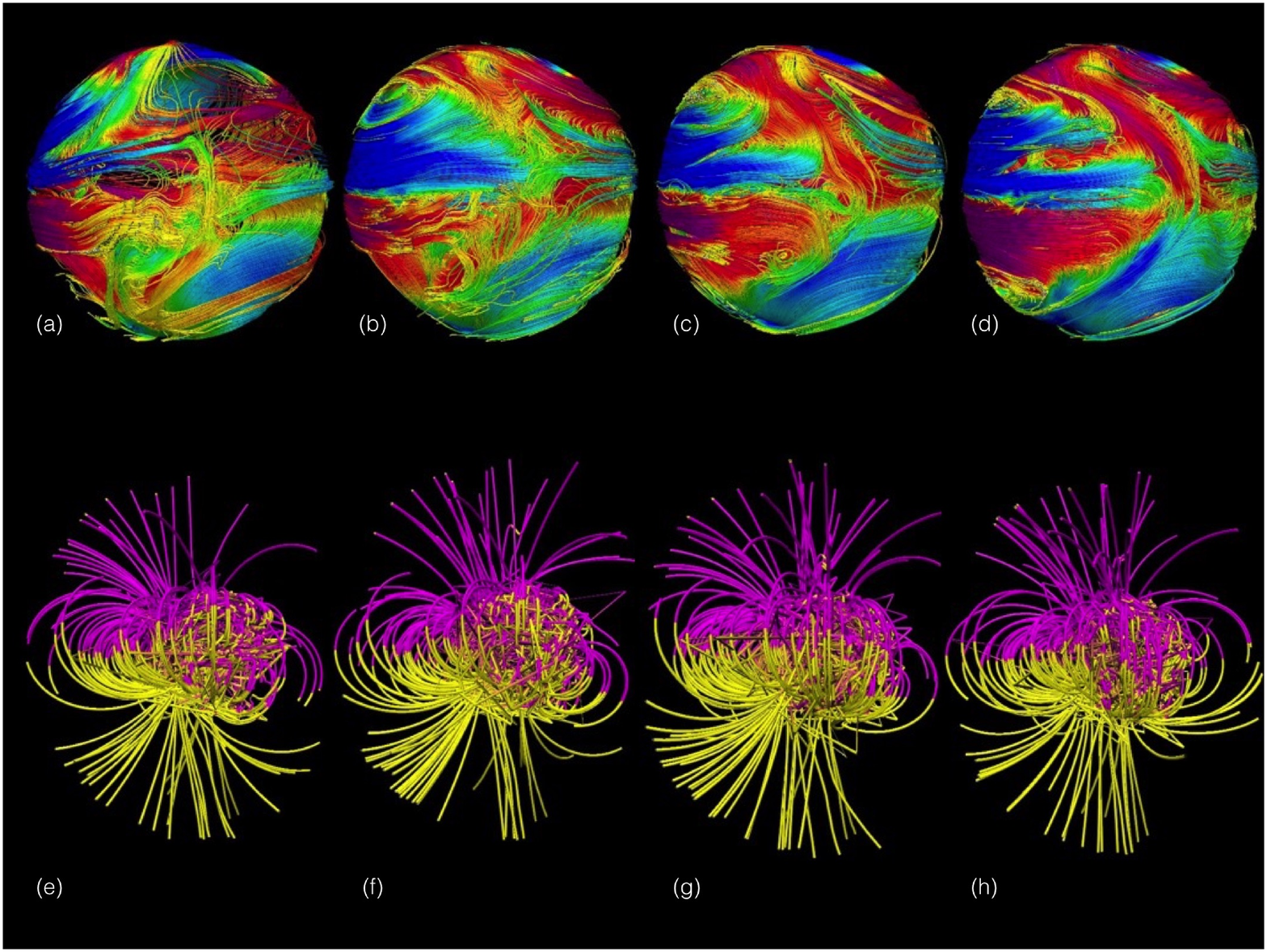}
\caption{Time snapshots of toroidal (azimuthal) magnetic field
  (looking onto the terminator) (a--d) and the radial magnetic field
  (e--h).}
\label{snapshots}
\end{figure}
The effect of the magnetic field on the atmospheric winds depends
  sensitively on the diffusivity profile and strength of the magnetic
  field, details of which can be found in \cite{rk14}. In the dynamo
  model presented here, the atmospheric winds are strongly coupled to
  the magnetic field and therefore, the winds become weaker and
  variable.  A time snapshot of magnetic field lines for the
``dynamo'' model, is shown in Fig.\ref{snapshots}. The top row shows
magnetic field lines looking onto the terminator\footnote{The
  terminator is the transition between day-night side, here we are
  referring to the terminator eastward of the sub-stellar point.},
color-coded by the azimuthal field strength, with blue positive and
magenta negative. Magnetic field is swept from the dayside, where
field and flow are strongly coupled, to the nightside, where much of
this field is dissipated. The collision between the strongly coupled
field on the dayside and the weakly coupled field on the nightside
leads to complex field topology and magnetic energy generation.
This interaction particularly generates strong latitudinal field at
the terminator, as can be seen in Fig. \ref{snapshots}d.

\begin{figure}
\centering
\includegraphics[width=6in]{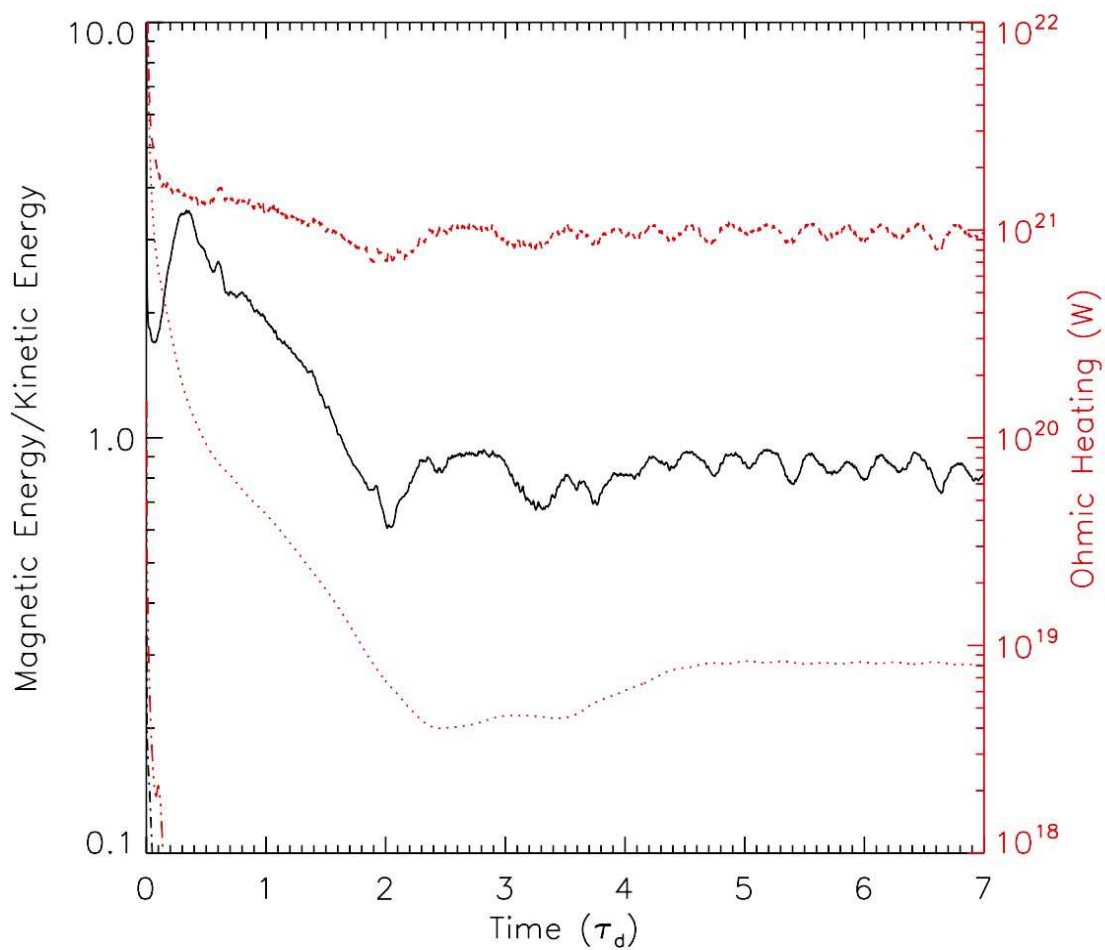}
\caption{Magnetic energy and Ohmic heating as a function of time,
  displayed in diffusion times, using the mean diffusivity
  $10^{11}$\,cm$^2$\,s$^{-1}$. The left hand axis shows the ratio of
  magnetic to kinetic energy (solid black line), clearly showing a
  dynamo as the magnetic energy is maintained against diffusion. The
  right hand axis shows Ohmic heating integrated below 10\,Bar (dotted
  line) and over the whole layer (dashed
  line). }
\label{magen}
\end{figure}

Investigation of (\ref{eq:induction}), shows that the VCD $\alpha$
effect is $\propto\Jvec\times\grad\eta$. The current, $\Jvec$, is
strongly correlated with vorticity, which tends to be strongest at the
day-night terminator and on the nightside \citep{rk14}. The
terminator is also where conductivity gradients are large, hence in
this region the necessary conditions for a dynamo are satisfied.
Specifically, we find that the radial component of the magnetic field
is regenerated predominantly from the azimuthal diffusivity gradient
$\sim \grad\eta_{\phi}J_\theta$ and the latitudinal component is regenerated predominantly from the radial
diffusivity gradient $\sim\grad\eta_rJ_{\phi}$. The azimuthal
component of the magnetic field is regenerated by \textit{both} the
typical $\Omega$ effect and by radial gradients in magnetic
diffusivity, $\sim\grad\eta_rJ_{\theta}$. The presence of a dynamo is
confirmed in Fig.~\ref{magen} which shows the ratio of the magnetic to
kinetic energy as a function of time for the ``dynamo model'' (solid
line) and the ``constant $\eta$'' model (dash-dot line-drops so
precipitously it can barely be seen in bottom left corner).

In the saturated state the magnetic energy generation is balanced with
Ohmic heating, which we also show in Fig.~\ref{magen}. If we compare
the Ohmic heating here to the values obtained in \cite{rk14} for a
3\,G imposed field (see their Table 1) we see that the Ohmic heating
here is equivalent to the Ohmic heating in a cooler model (between
M6b3 and M7b3). Therefore, we conclude that the presence of a
horizontally varying conductivity and the dynamo it produces, results
in slightly \textit{lower} overall Ohmic heating than one would expect
from a model which does not consider a horizontally varying
conductivity. This is likely because magnetic energy is maintaining
the VCD, rather than being dissipated.

The bottom row of Fig.~\ref{snapshots} shows the extrapolated field
lines (out to 2R$_p$) for this model. There we see that the magnetic
field is very asymmetric, with poloidal field concentrated
predominantly on the dayside of the planet. Near the equator the
surface poloidal field (defined as $B_p=\sqrt{B_{\theta}^2+B_{r}^2}$)
is $\sim$15\,G on the dayside of the planet and 7\,G on the nightside of
the planet. These values are 16\,G/8\,G for the ``imposed+dynamo''
model and $\sim$1\,G on both the day and night side for the
``imposed+constant'' model. The inclusion of conductivity variations
significantly increases the surface planetary field strength and leads
to a highly asymmetric field. Therefore, unless the internal,
convectively driven, magnetic field is particularly strong (in excess
of 15\,G at the surface), the surface planetary magnetic field is
likely dominated by the magnetic field generated in the atmosphere, at
least in hotter hot Jupiters.

The asymmetry in the field persists to $2R_p$ where about 65$\% $of
the magnetic energy is found in the dipole component, 25$\%$ is in the
$l=2$, $m=1$ component and 10$\%$ in the $l=3$, $m=2$ component,
although by $4R_p$ the dipole component represents 95\% of the total
energy. However, these percentages fluctuate significantly in
time. Such a complex field structure in space and time likely affects
the SPMI expected in these close-in systems
\citep{cuntz00,ip04,sbm15}.
  
\section{Analytic models of dynamo behavior in a hot Jupiter} 

The flow profile and conductivity variations in hot Jupiters are
relatively simple, therefore, we attempt to solve this system
analytically. While the diffusivity and initial velocity vary on a
large scale in a hot Jupiter atmosphere, the magnetic energy is
generated in a narrow region near the terminator. In that region the
length scale of velocity and diffusivity variations is small.
Therefore, we apply the standard technique of multiple scales from
homogenization theory \citep{homog} in 3D cartesian coordinates.  This
is a method for deriving an equation for the large scale variations in
the magnetic field by averaging over the periodic small scale
variations. We assume that the conductivity and velocity vary on
  a small spatial scale defined by a large wavenumber $k$ and define
  $\bX=(X,Y,Z)=k\bx$ and all functions must be periodic in $\bX$. We
  then look for a solution of the form
  \begin{equation}\label{eq:H1}
    \bB = \bB_0(t,\bx)+k^{-1}\bB_1(t,\bx,\bX)+k^{-2}\bB_2(t,\bx,\bX)+\cdots.
  \end{equation}
  When we substituting (\ref{eq:H1}) into (\ref{eq:induction}) we get
  a series of equation at each order in $k$. The leading order
  equation, $O\left(k^1\right)$, gives an equation for $\bB_1$ in
  terms of $\bB_0$ and the $O\left(k^0\right)$ equation, after
  averaging over the periodic cell, gives an equation for $\bB_0$.
  The $\bB_0$ equation is
  \begin{eqnarray}\label{E1}
    \nabla' \times \left[ (\eta (\nabla\times \bB_0 + \nabla' \times  \bB_1) -\bu \times \bB_0\right] &=&0,
  \end{eqnarray}
  where $\nabla'=(\partial_X,\partial_Y,\partial_Z)$. This is to be
  contrasted with (4) in \cite{pet16}, where there is no spatial
  variation in $\eta$ and a time derivative is included. We then write
  $\eta=\eta_0(1+\delta\eta')$, where $\eta_0$ is constant, and
  $\eta'$ varies between $\pm 1$, thus $\delta$ controls the strength
  of the variation in conductivity. Writing
  $\bB_1 = \sum_{k=0} \delta^k \bC_k$ and Taylor expanding
  $(1+\delta\eta')^{-1}$ we get a series of Poisson equations for
  $C_k$ which can be easily solved to give $\bB_1$ in terms of $\bB_0$
  and $(\nabla \times \bB_0)$. This expansion is convergent for
  $\delta<1$, which we also expect on physical grounds since this
  corresponds to positive diffusivity everywhere.

We choose profiles similar to those found in hot Jupiters.
Assuming $\xh$, $\yh$ and $\zh$ correspond to the azimuthal,
latitudinal and radial directions, we write
\begin{equation}
  \eta'=\cos Z \sin X \sin Y,
\end{equation}
and
\begin{equation}
  \bu  =\sin Z \left( \xh U \sin X \cos Y + \yh V \cos X \sin Y \right)
\end{equation}
Here $U$ and $V$ represent the azimuthal and meridional velocity
amplitude, respectively. As discussed in Section 3 the winds in the
hot Jupiter atmosphere are largely two-dimensional and are reasonably
described by (11).  Using the method described above we solve
  for $\bB_1$ to $O(\delta^4)$ and the $\bB_0$ equation becomes
\begin{equation}\label{eq:E0}
  \frac{\partial \BZ}{\partial t}  
  =
  \eta_0\left[1-\frac{\delta^2}{12}-\frac{\delta^4}{72}\right]\nabla^2 \BZ
  +\eta_0\frac{\delta}{24}\left[1+\frac{\delta^2}{6}+\frac{233881}{3852288}\delta^4\right] \nabla
  \times \BZ' + O(\delta^6),
\end{equation}
where $\BZ'=-VB_x\xh+UB_y\yh+(U-V)B_z\zh.$ The
  $\nabla \times \BZ'$ term gives rise to the dynamo effect. When
$UV>V^2$ (which is the appropriate case for hot Jupiters), the large
scale magnetic field is unstable in the $\xh$ direction and the
minimum critical Reynolds number, defined as $UL/\eta$ where L is the length
  scale of the large scale magnetic field, is\footnote{This stability condition
  is dependent on the exact diffusivity and velocity profile. While
  many profiles give instability, many do not and we are still in
  the process of finding a generalized solution to the conditions
  for instability.}
\begin{equation}
  R_{mc} > \frac{\mbox{$\scriptstyle\sqrt{1+\sqrt{2}}$}\,(6-\delta^2)(12+\delta^2)}{\delta^2\left(6 + \delta^2 + \frac{233881}{642048}\delta^4\right)}.
\end{equation}

\begin{figure}
  \centering\label{fig:Rm}
  \includegraphics{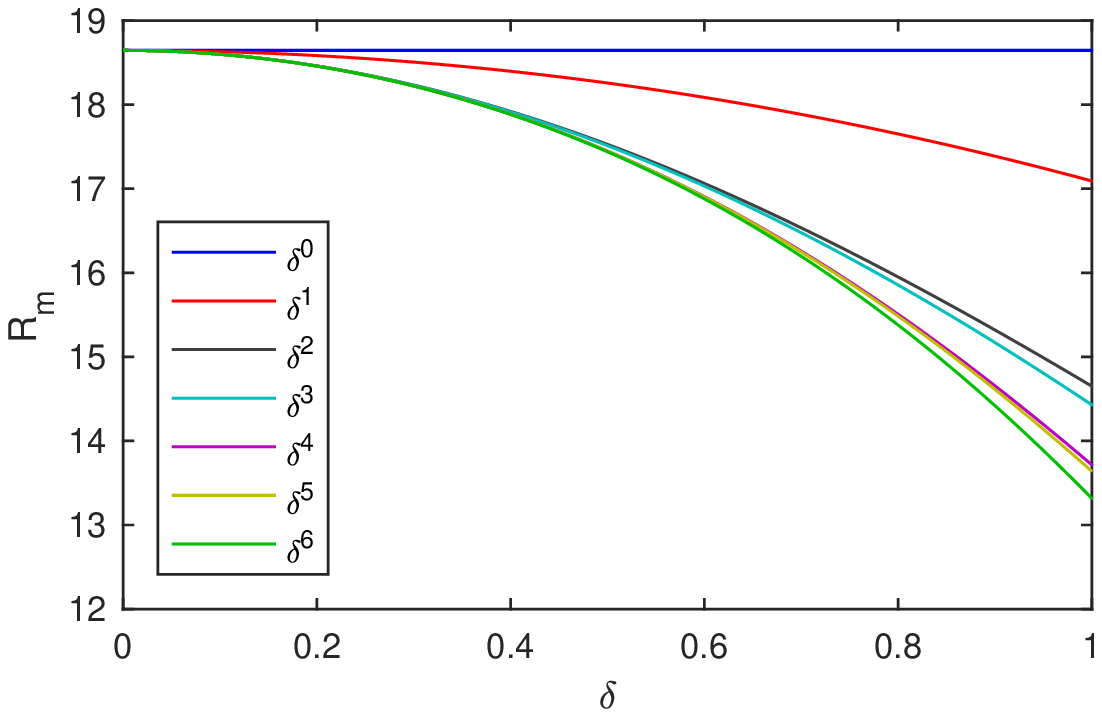}
  \caption{Scaled Critical Magnetic Reynolds number $\delta^2R_{mc}$ as a function of $\delta$
    for different accuracies of $\bB_1$ solution. }
\end{figure}
In hot Jupiters, the day-night diffusivity,
$\eta_{max}/\eta_{min}=(1+\delta)/(1-\delta)$ varies between
$\sim$10$^1$--$10^{4}$, which corresponds to a $\delta$ between
0.9--0.999. Figure~\ref{fig:Rm} shows the approximate
  convergence of $R_{mc}(\delta)$ as $\delta$ approaches 1.  If we take
$\delta=0.9$ the $R_{mc}$ needed for instability is $\sim$14.  Using
the length scale over which magnetic energy is generated at the
terminator ($\sim$10$^{10}$\,cm) and a typical velocity of
\mbox{$\sim$10$^4$\,cm\,s$^{-1}$}, we conclude that in order for a hot
Jupiter atmosphere to host a dynamo, the nightside magnetic
diffusivity must be
$\lesssim$10$^{12}$--$10^{13}$\,cm$^{2}$\,s$^{-1}$, or a temperature
of roughly 1400\,K on the \textit{nightside} of the planet. We expect
the estimate for the diffusivity could vary by about an order of
magnitude due to variations in metallicity and the inclusion of a
temperature dependent diffusivity. Therefore, we conclude that a VCD
only occurs in the hotter hot Jupiter atmospheres.

\section{Discussion}
Using numerical simulations coupled with analytic results we have
shown that hot Jupiters with night side temperatures which are above
$\sim$1400\,K could host dynamos driven by spatial conductivity
variations due to asymmetric heating from the host star. Lower
  temperatures and weak day-night temperature differentials do not
  produce dynamos. This is remarkable, not just because the dynamo is
driven by conductivity variations (as has been shown previously), but
also because it is maintained in a stably-stratified, thin atmosphere.
The inclusion of horizontal variations in conductivity reduces Ohmic
heating compared to a similar temperature object with no such
variations. However, it is hard to make a direct comparison because
of the large conductivity variations. To really investigate Ohmic
heating, the analysis presented here will have to be done for a host
of planetary temperatures and day-night temperature differences,
something which is currently underway. Moreover, one needs to include
the recovered Ohmic heating in a planetary evolution model to make a
more concrete statement about the viability of Ohmic heating in
explaining hot Jupiter radii. Finally, we will have to consider the
interaction of the atmospheric field with the convectively generated
field.

Whatever the deep seated field, it will be subject to interaction with
the atmospheric winds and the variable conductivity in the atmosphere,
both of which affect the overall field strength and geometry. We find
that, unless the deep seated dynamo magnetic field is unreasonably
strong, the surface planetary magnetic field strength is dominated by
the \textit{induced} field, particularly on the dayside of the planet.
We also find that the magnetic field geometry is asymmetric, with
dayside fields approximatley two times larger than their nightsides
(dependent on the day-night temperature difference). Furthermore, we
find that the energy in the dipole component of the magnetic field
varies substantially in time. All of these factors affect star-planet
magnetic interactions \citep{sbm15} and the inferences we make from
such interactions \citep{vidotto10b}.

While we have made progress by including a spatially dependent
conductivity, we have yet to consider a \textit{temperature-dependent}
conductivity. We expect this will play an important role, possibly
leading to the instability proposed by \cite{menou12b} and likely
increasing the overall Ohmic heating and altering still further the
magnetic structure. Such simulations will be the subject of a
forthcoming paper.

\bibliographystyle{apj}

%\bibliography{hjm}

\acknowledgments 
Support for this research was provided by NASA grant NNX13AG80G to
T.Rogers. Computing was carried out on Pleiades at NASA Ames.
T. Rogers thanks Graeme Sarson, Gary Glatzmaier and Chris Jones for
useful conversations leading to the development of this manuscript.

\end{document}